# COUPLED BUNCH INSTABILITIES GROWTH IN THE FERMILAB BOOSTER DURING ACCELERATION CYCLE *

C. M. Bhat[†] and N. Eddy, Fermilab, Batavia IL, U.S.A.

## Abstract

Currently, Fermilab Booster accelerates ~4.5E12 protons per pulse (ppp) in 81 bunches from 400 MeV to 8 GeV at 15 Hz to provide beam to multiple HEP experiments and is being upgraded to handle higher beam intensity >6.7E12 ppp at a repetition rate of 20 Hz. In the current mode of operation, we control longitudinal coupled bunch instabilities (LCBI) using combination of passive and active dampers. The issues with LCBI are expected to worsen at higher beam intensities. Here, we would like to investigate at what time in the beam cycle a particular mode is going to originate and how much it contributes at a different time of the acceleration cycle using the collected wall current monitor data from injection to extraction. Also, measure growth rates as a function of beam intensity for some dominant modes. Finally, this paper presents results from the observed LCBI analysis for the highest available beam intensities, unexpected findings, and future plans to mitigate the LCBI at higher beam intensities.

## INTRODUCTION

The Fermilab Booster is one of the oldest RCS in the world which cycles at 15 Hz rate [1] and providing 8 GeV beam to HEP programs for the last 50 years. Until 2016 the beam in the Booster was accelerated with low duty factor, e.g., during the early part of the Main Ring fixed target HEP era it provided ~ 2E12 ppp with an average repetition rate slightly more than 0.5 Hz out of 15 Hz and with lower beam brightness. In early 1980s, as we entered the collider era along with fixed target HEP programs, the demand on the beam intensity from the Booster per cycle as well as the beam delivery repetition rate increased. All these periods, the longitudinal coupled bunch instability was troublesome in the Booster and, was being addressed and controlled [3-8]. In 2012, Fermilab entered the intensity frontier era PIP [2] and by the end of this decade it will be in PIP-II era [9]. Currently, we are providing > 4.3E12 ppp at 15 Hz rate from the Booster and, >700 kW proton beam power on the NuMI neutrino target and at the same time beam to multiple Booster neutrino experiments. Between now and when the PIP-II comes online, we have a near-term goal to increase the beam output power per Booster cycle by about 25%. PIP-II calls for a further increase in beam intensity by ~50% and a cycle rate of 20 Hz. In this regard the coupled bunch instability needs to be revisited.

Since the start of the Booster, the beam has been injected into the Booster without any RF buckets opened. After the completion of the injection the beam is captured slowly in the RF buckets with harmonic number $h$=84. So, it is expected that the bunches are well matched to the RF buckets from the start of their formation with minimal longitudinal instabilities. However, during the acceleration from injection energy to 8 GeV beam encounters transition crossing at $\gamma$ =5.47 with normal transition phase jump and we have observed significant longitudinal emittance growth only after transition energy. This emittance growth is related to longitudinal coupled bunch instability observed after transition crossing [4, see, K.C. Harkey, Ph.D. Thesis]. Since then, multiple upgrades were made to add passive [5] and active [3, 6, 8] dampers to mitigate LCBI in the beam. In all these cases it was enough to enable the active dampers only after the transition crossing and monitor the coupled bunch instability in the extracted beam using a wall current monitor (WCM) in the 8 GeV transfer line. During PIP-II era, 800 MeV beam will be injected directly into the Booster RF buckets and by the end of the injection process the beam distribution is expected to match well with the RF buckets. Nevertheless, the bunch intensity will be significantly higher than that in the current operation. As a result of this, it is of interest to investigate the evolution of LCBI in Booster for high intensity beam ahead of PIP-II and develop mitigation strategy.

## ASPECTS OF DIPOLE LCBI

Theory of longitudinal coupled bunch instability and stability criterian are one of the widely discussed subjects in beam physics [10-12]. An application of the theory to study and control the coupled bunch dipole mode instability in the Fermilab Booster was carried out by D. P. McGinnis [3] in early 1990s. Here, we highlight some key aspects of the theory in the context of studies carried out for the intensity upgrade in the Booster.

Circulating bunches of charged particle in a synchrotron give rise to longitudinal as well as transverse wake fields because of machine impedance including higher-order modes of RF cavities. The longitudinal wake field which is not decayed between arrival of consecutive bunches in the ring will be responsible for longitudinal coupling between bunches and is an important source of instability in longitudinal plane. The wake fields can potentially cause both bunch displacement as well as distortion and help growing exponentially during successive passage in the synchrotron. The particles in bunches execute synchrotron oscillations due to RF force. So, the displaced bunches will also experience this additional force in the RF bucket which adds up for further emittance dilution and possibly beam loss during the beam acceleration. There are other sources of bunch oscillations which may not be due to wake field effects. Such examples are energy and phase mismatch during bucket-to-bucket bunch transfer from one accelerator to the other or in the case of space charge dominated beam, immediately after the transition crossing with RF phase jump from $\phi$ to $\pi$-$\phi$. Such bunch oscillations are generally not considered as longitudinal beam instability. However, they can induce LCBI. In any case, the line densities of such oscillating bunches can be decomposed into two parts *viz.*, stationary distribution and additional charge density oscillating in synchrotron phase space. The oscillation part is approximately sinusoidal standing-wave with $m$ azimuthal nodes. (these are intra-bunch synchrotron oscillation modes). For example, a single bunch can perform rigid dipole oscillations due to phase mismatch which corresponds to mode $m$=1 and in the case of RF voltage mismatch one can expect quadrupole oscillations with $m$=2 (also called breathing mode). A nearly triangular shape distribution of particles in $(dE, \Delta t)$ phase-space in a RF bucket would lead to sextupole oscillations with m=3 and so on. Sometimes multiple modes can occur at the same time. To drive higher azimuthal modes, longitudinal impedance of higher frequencies are needed. The standard formalism is to transform the bunch displacements into Fourier modes. The modes can be identified by the lines that can be seen in bunch Fourier spectrum [10-12] with

$$f_p = (n + pM)f_0 \pm |m|f_s$$
$$\text{with } p = 0, \pm 1, \pm 2, \ldots \pm \infty \text{ and}$$
$$|m| = 0,1,2,\ldots \infty \quad (1)$$

_________________
* Work supported by Fermi Research Alliance, LLC under Contract No. De-AC02-07CH11359 with the United States Department of Energy
† cbhat@fnal.gov.

$f_0$ is revolution frequency of the bunches, $f_s$ is synchrotron frequency, $M$ is number of bunches, generally same as RF harmonic number $h$, $n$ is mode number for LCBI. In the presence of LCBI the spectrum consists of additional signal at those rotation harmonics corresponding to $(n + pM)f_0$ which is characteristic of coupled bunch mode $n$. $p$ represents band number; negative and positive values of $p$ correspond to lower and upper bands, respectively. As mentioned earlier, a single bunch can perform dipole, quadrupole, and higher ordered mode oscillations. In addition to this, all $M$ bunches may oscillate with same mode $m$ and with same amplitude but, with constant phase advance of $2\pi n/M$ between bunches due to coupling. For $M$ identical bunches there are $M$ coupled bunch modes of oscillations and the mode number $n$ takes the values $0, 1, 2, \ldots, M - 1$. Generally, a mixture of different oscillation modes $m$ and $n$ is observed. Selection of a specific mode of interest for observation or mitigation can be done by a discrete Fourier transform (DFT) or by measuring the beam signal within a limited bandwidth and supress it as needed. In the current measurements and analysis, we use the former method.

As mentioned earlier, the phase-space distribution $\psi$ of a bunch can be separated into stationary $\psi_0$ and small oscillations about the stationary distribution $\psi_1$ as follows [10-12],

$$\psi(r, \phi, t) = \psi_0(r, \phi) + \psi_1(r, \phi, t)$$
$$= \psi_0(r, \phi) + \sum_{m=1}^{\infty} R_m(r) e^{-im\phi} e^{-i(mf_s + \Delta f_m)t} \quad (2)$$

$\Delta f_m$ is coherent frequency shift [10]. In the above equation the quantities $r$ and $\phi$ are radius and angle, respectively, in the longitudinal phase space that have been normalized to make the unperturbed trajectories circular. The variable $t$ represents location of the test particle [12, chapter 1]. $R_m(r)$ is a function corresponding to $m$th azimuthal mode. The distribution functions for other bunches are identical to Eq. 2 except that the modes are multiplied by phase advance of $2\pi n/M$ per bunch. The complete distribution function for bunches is a superposition of $M$ function of the type given by Eq. 2. Thus, complete distribution function is given by,

$$\psi(r, \phi, t, k) = \psi_0(r, \phi)$$
$$+ \sum_{n=0}^{M-1} \sum_{m=1}^{\infty} B_n R_p(r) e^{i2\pi nk/M} e^{-im\phi} e^{-i(mf_s + \Delta f_m)t} \quad (3)$$

where $k$ is the bunch index. $B_n$ is the relative strength of the various azimuthal modes. As shown in the ref. 3, by measuring the average phase error $<\phi_k>$ for $k$th bunch and performing DFT one can extract $B_n$ for each mode. Notice that this method of extraction is exact for dipole modes only, which are dominant modes of LCBI in most of the cases at hand.

We start with measurement of total distribution function $\psi$, given by Eq. 3, which is snap-shot single turn line charge distribution data as a function of time for circulating train of $M$ bunches. This can be obtained using a resistive wall current monitor (WCM) of enough band width and a high-speed oscilloscope which gives enough resolution for each bunch. Selecting the first bunch is somewhat arbitrary if there is no single turn marker like a notch in the beam. If there is notch in the beam, then the first bunch from the notch can be used as a reference bunch. The actual centroid of $k$th bunch $bc_k$ can be determined using a relation like,

$$bc_k = \frac{\int t\psi_k(t)dt}{\int \psi_k(t)dt} \quad (4)$$

where $\psi_k(t)$ is distribution function (or WCM data) for $k$th bunch. The extent of each bunch is one turn. As we have explained earlier, the relative change in spacing of adjacent bunch centroids is due coupled bunch mode oscillations which is a combination of unperturbed centroid measured with respect to the RF wave $\phi_0$, linear increase of phase by $2\pi n/M$ at a given RF frequency $\phi_1$, due to slewing of the RF frequency during beam acceleration $\phi_2$, and other negligibly small higher order terms. Thus, the phase can be written as,

$$<\phi_k>_{fit} = \phi_0 + \phi_1 k + \phi_2 k^2 + \cdots \quad (5)$$

The phase errors $\Delta \phi_k$ due to the coupled bunch motion are obtained by,

$$\Delta \phi_k = bc_k - <\phi_k>_{fit} \quad (6)$$

From Eq. 6 one can extract the mode strengths by Fourier analysis. In the case of Booster with 84 bunches one expects 84 modes of oscillations. The Fourier spectrum obtained from $\Delta \phi_k$ on one turn beam data displays 84 lines with each line corresponding to one value of mode number

$n$. Furthermore, magnitude of mode $n$ (=0-41) is exactly same as that for $83 - n$ [12].

## MEASUREMENTS AND DATA ANALYSIS

The distribution function for the circulating beam in the Booster is obtained using a 6 GHz bandwidth wall current monitor (WCM). A Tektronix, TDS7154B Digital Phosphor Oscilloscope of type 1.5GHz 20GS/s is used in the Booster to record the WCM data. At the same time, RF waveform data is collected in synchronized with WCM signal. Due to limited memory size of the scope, about 300 traces of data with 67 turn delay which covers entire beam cycle are collected. The RF frequency in the Booster swings from about 37.92 MHz to 52.81 MHz during the beam acceleration from 400 MeV to 8 GeV, correspondingly, the Booster beam revolution varies from 2.22 μs to 1.59 μs from injection to extraction. The WCM data for each trace spans at least one Booster turn. During the beam cycle 81 buckets out of 84 buckets are populated leaving three buckets empty which are used as kicker gap at the time of beam extraction. A typical mountain range of the WCM data is shown in Fig. 1. The beam between notches in a trace represents total beam in Booster. A Mathcad program is developed to perform the rest of the data analysis following the procedure explained in the previous section.

Since the beginning of PIP era, we have been upgrading our longitudinal dampers from analogue to digital to suppress LCBI for modes 1,2, 45-53 which have become dominant modes of beam instability. We use all our twenty RF stations to control modes 1 and 2 [13]. During early part of the beam acceleration where time derivative of the frequency is very large, it uses almost all available RF voltage of ~1 MV and as beam crosses the transition energy at about 17 ms in the cycle the required RF voltage for further beam acceleration will go down nearly by 10%. The mode 1, 2 damper uses this excess of RF power and is turned on only after the transition crossing. The modes 45-53 (same as 30-38) damper uses a dedicated 80 MHz broadband rf cavity [8]. One can turn ON and OFF the dampers on individual modes separately. The beam data have been taken under varieties of damper configurations and beam intensities.

### An illustration for extracting LCBI Mode Strength from a Single Turn Data

Figure 2 illustrates an example of extracting the coupled bunch mode strengths for dipole mode oscillations using single turn WCM data for 1.8E12ppp in Booster with all longitudinal dampers turned OFF. The extent of each bunch in a batch of 81 (84 buckets) is established using the RF waveform as a reference. An example of such RF waveform along with the corresponding WCM data for all 81 bunches is shown in Fig. 2(a). Figures 2(b) and 2(c) are for 81 bunches shown in the form of mountain range and contour plots which helps one to observe coupled bunch motion visually. The reference RF waves used in establishing bunch extents are shown in Fig. 2(d). Extracted bunch centroid as a function of bunch number is shown in Fig. 2(e). The measured phase error according to Eq. 6 in units of RF degrees for each of the bunches and its DFT are shown in Fig. 2(f) and 2(g), respectively.

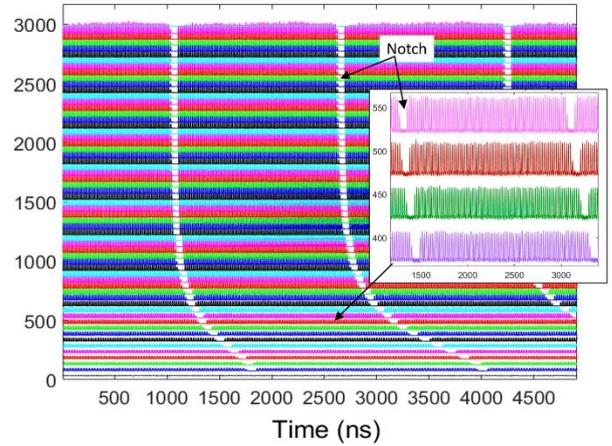

Figure 1: Mountain Range of the Booster WCM data from injection to extraction. The x-axis scale is time around the ring. The scale on the y-axis is also time with 355 turn delay between traces. This large turn delay is selected only for clarity.

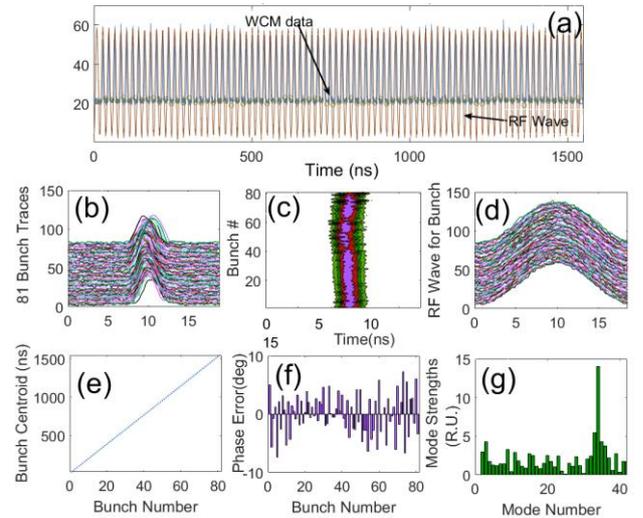

Figure 2: A sample of (a) Line-charge distribution with RF wave. The details on (b) to (g) are explained in the text.

### LCBI Mode Strengths through the Beam Cycle and Growth Rate for some Dominant Modes

Here we present the LCBI results for various beam intensities in the range from 1.2E12 ppp to 3.6 E12 ppp with longitudinal dampers turned OFF. Figure 3(a) and (b) illustrate LCBI analysis for turn #9581 and turn # 19631, slightly below the transition energy (transition energy is around turn number 9700) and near extraction well above transition energy, respectively, for beam intensity 3.6E12ppp. It is quite clear from Fig. 3(b) mountain range plot (left) that there is significant amount of dilution in longitudinal emittance. Also, we observe considerable amount of beam loss before beam extraction. Before transition to after transition the coupled bunch instability strength differ by more than a factor of twenty for several modes at this intensity.

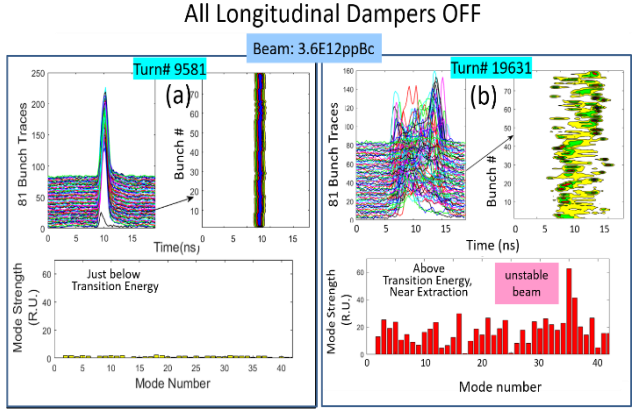

Figure 3: Extraction of Booster dipole LCBI strengths (a) below transition energy and (b) very near the extraction energy (after transition energy).

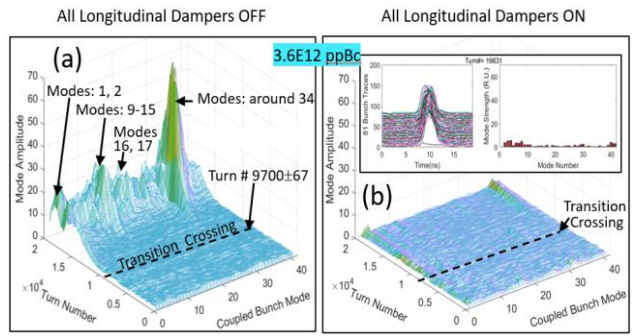

Figure 4: Surface plots for extracted LCBI strengths from injection to extraction in the case of 3.6E12 ppp with (a) all longitudinal dampers OFF, (b) all dampers ON. The insets are, respectively, mountain range for all 81 bunches (left) and measured mode strengths with dampers ON (right).

Figure 4(a) and 4(b) illustrates typical surface plots of the measured LCBI strengths from injection to extraction for all modes from 0-41 for beam intensity of 3.6E12 ppp with all active longitudinal damper turned OFF and turned ON, respectively. In Fig. 4(a) an approximate turn number corresponding to transition energy during the beam acceleration is also shown just for a comparison of the mode strengths before and after the transition energy. The data shows that below transition crossing very little LCBI is observed for any mode. On the other hand, above transition energy almost all modes showed up with considerable strengths. Among them, modes 1,2, some modes around 12, 17 and 34 (same as 49) showed up with noticeable strengths.

Figure 5 shows growth rates for some dominant modes as a function of beam intensity. The data shows that modes 33, 34 and 35 (50, 49 and 48), have significantly higher growth rate as compared with others. Furthermore, for beam intensities >2E12 ppp without longitudinal dampers turned ON, we have observed significant longitudinal beam emittance growth as well as noticeable loss during transport to the downstream experimental facilities of the Booster. With all active dampers turned ON (and phased-in), we were able to mitigate LCBI instability as shown in surface plot in Fig. 4(b). The bunches are quite centered as shown in left inset of Fig. 4(b) (compared with similar mountain range of bunches shown in Fig. 3(b) with dampers OFF).

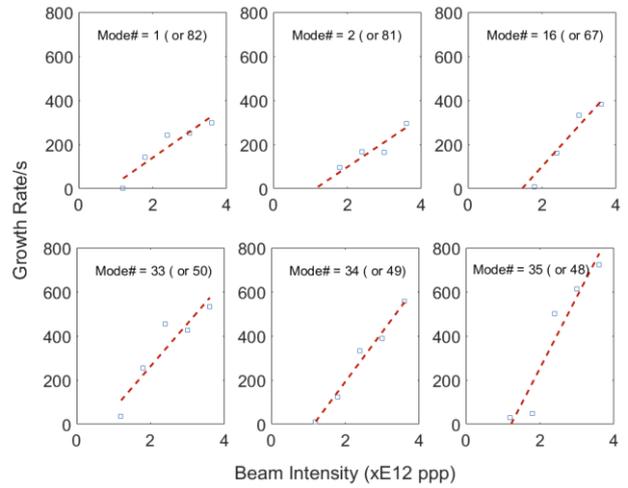

Figure 5: LCBI growth rate as a function of beam intensity for mode 1,2, 16, 33, 34 and 35.

*Operational and higher intensity cases with dampers turned ON*

The measurements showed that in the current mode of operation the highest beam intensity that can be accelerated in the Booster without any loss pass transition energy is about 3.6E12 ppp. For intensities over 3.6E12 ppp we must have the digital dampers turned ON and adjust the phase and magnitude as needed to mitigate the LCBI. Figure 6 shows surface plot for measured LCBI strengths for all modes with dampers for 4.5E12 ppp which is slightly larger than PIP design beam intensity. The inset Fig. 6(a) illustrates the results from a typical longitudinal emittance measurement after beam capture at 400 MeV (top) and very close to the extraction at 8 GeV (bottom). The measured beam longitudinal emittances $\varepsilon_L(95\%) \sim 0.07 \pm 0.01$ eVs and $\sim 0.11 \pm 0.01$ eVs, respectively, at the end of the capture and near to the extraction in the Booster. This beam is acceptable by the facilities downstream of the Booster. Studies showed that about 10% smaller emittances can be reached by improving capture process at 400 MeV [13]. Average injection to extraction efficiency was about 94% at these beam intensities. The inset Fig. 6(b) shows a time evolution of LCBI strength for mode # 39 (same as mode 44) averaged over three data samples. This is an indication that this mode could be a concern at higher intensities and needs better damping.

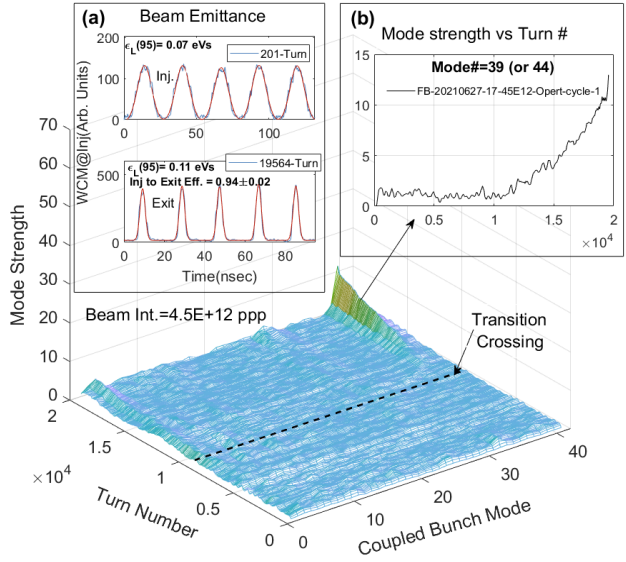

Figure 6: Surface plot showing all modes for the case of 4.5E12 ppp at extraction with longitudinal dampers turned ON. The insets (a) and (b) display typical beam emittance measurements and mode strengths for mode 39.

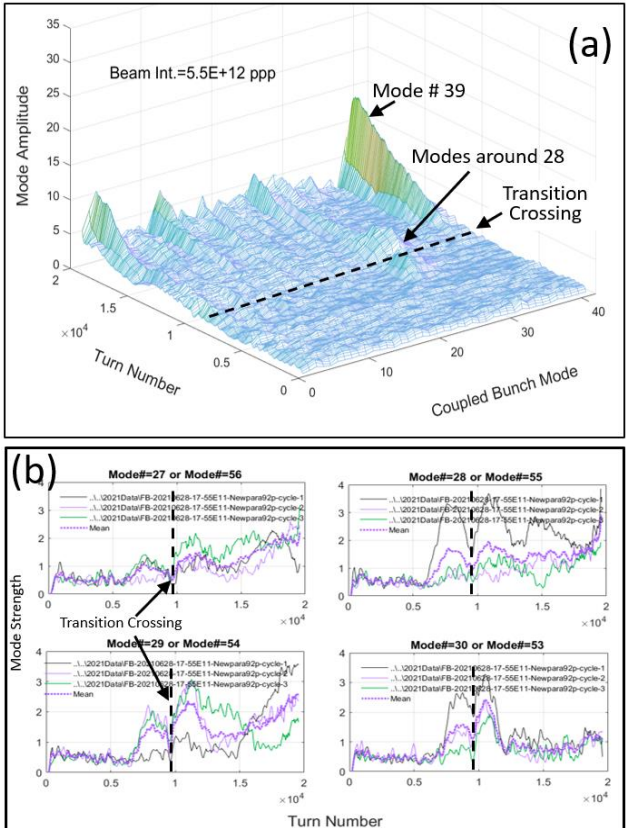

Figure 7: (a) Surface plot showing all modes for the case of 5.5E12 ppp at extraction with longitudinal dampers turned ON. (b) Some typical modes, 27-30, showing up even before the transition crossing.

Figure 7(a) shows similar surface plot for the LCBI strengths for all modes for the highest beam intensity studied here. Injection to extraction efficiency for this case was about 92%. Multiple modes started showing up here with non-negligible LCBI strengths. As indicated in Fig. 7(a) and in Fig 7(b), we also start seeing modes near 28 start showing up even before transition crossing. The reasons for the mode strengths going down before transition energy by themselves for the cases shown in Fig. 7(b) is not well understood and is part of our future studies. By optimizing the capture process, we had longitudinal emittances $\varepsilon_L(95\%) \sim 0.06 \pm 0.01$ eVs at the end of beam capture and $\sim 0.09 \pm 0.01$ eVs near extraction, respectively.

The current study with high intensities >5E12ppp clearly shows that a) multiple mode not damped with the existing digital dampers start growing even before the transition energy and b) some of the modes which are damped using existing dampers at beam intensities ~5E12 ppp start popping up after the transition crossing. Therefore, for PIP-II era Booster beam intensities ~ 6.7E12 ppp at extraction we need further upgrades to our longitudinal dampers to mitigate modes around 12, 16 and 39 in addition to other modes. In this regard, we are planning to build dedicated longitudinal damper systems including new rf cavities to supress modes 1, 2 and modes around 16 and 39 in a few years' time. Nevertheless, these problems may not preclude us to achieve our intermediate goal of providing 900 kW beam power on the NuMI neutrino target with the existing longitudinal dampers in Booster.

## ACKNOWLEDGEMENTS

We would like to thank W. Pellico, C. Y. Tan and S. J. Chaurize for many discussions. We also would like to thank MCR crew for their help during these measurements.